\documentclass{iopart}

\usepackage{graphicx}
\usepackage{enumerate}
\usepackage{iopams}
\usepackage{color}
\usepackage{bm}
\def\n{\mathbf{n}}

\begin{document}
 
\title[]{Ground-state properties of the spin-$\frac{1}{2}$ antiferromagnetic Heisenberg model on the triangular lattice: A variational study based on entangled-plaquette states}
\author{F Mezzacapo and J I Cirac}
\address{Max-Planck-Institut f\"ur Quantenoptik, Hans-Kopfermann-Str.1, D-85748, Garching, Germany}
\ead{fabio.mezzacapo@mpq.mpg.de}

\begin{abstract}
We study, on the basis of the general entangled-plaquette  variational ansatz, the ground-state properties of the spin-$\frac{1}{2}$ antiferromagnetic Heisenberg model on the triangular lattice. 
Our  numerical estimates are  in good agreement with available exact  results and comparable, for large system sizes, to those computed via the best alternative numerical approaches, or by means of variational schemes based on  specific  (i.e., incorporating problem dependent terms) trial wave functions.
The extrapolation to the thermodynamic limit of our results  for  lattices comprising up to $N=324$ spins yields  an upper bound of the  ground-state energy per site (in units of the exchange coupling) of $-0.5458(2)$ [$-0.4074(1)$ for the XX model], while the estimated infinite-lattice order parameter is $0.3178(5)$ (i.e., approximately $64\%$ of the classical value).
\end{abstract}

\pacs{02.70.Ss, 05.50+q}
 

\section{Introduction}

In the last decades quantum Monte Carlo (QMC) \cite{qmc} has emerged as one of the most powerful numerical tools to study the ground-state (GS) properties of quantum spin systems on a lattice. It has been successfully applied, for example, to determine quantities of crucial interest such as the energy and the order parameter of the two dimensional (2D) antiferromagnetic Heisenberg model (AHM) on the square lattice \cite{sse} and, more generally, can be considered the method of choice to investigate unfrustrated (bosonic) models in any dimension.

The situation is more complicated when the geometry of the system is frustrated as in the triangular or kagom\'e  lattice. In this case QMC techniques based on imaginary time projection, which are ``exact" for unfrustrated systems, suffer from a sign-problem  and approximations or alternative techniques  have to be employed. The density matrix renormalization group (DMRG) \cite{dmrg} method, although not affected by sign instability, yields very accurate results only for one dimensional (1D)  spin systems, being instead hardly applicable to general 2D problems due to the unfavorable scaling of the computer resources needed with the system size \cite{dmrgsize}. 

Projected-entangled pair states (PEPS) \cite{peps}, the natural extension  to higher spatial dimensions of the matrix product states  variational family, which constitutes the foundation of DMRG, can efficiently approximate GS's of local Hamiltonians \cite{has}, and have been successfully employed to simulate frustrated and bipartite quantum spin systems \cite{peps1, peps1a, bauer, orusetal, orus}. Their applicability, however, is restricted by the   disadvantageous scaling of the computational cost to 2D models with open boundary conditions. Different approaches, which have been proposed, and applied in 2D \cite{vidalmera, vidalmera1, Levin, Weng, LevinWen, vidal2, vidal2a, vidal2b},  suffer  from similar limitations  and seem  hardly usable for large systems [possibly three dimensional (3D) ones] and periodic boundary conditions (PBC). The idea of combining DMRG and Monte Carlo \cite{SV,SBS} has opened the opportunity to design new variational algorithms for Tensor-Network based ansatze \cite{SBS1, Snew}. The applicability of one of them, based on  string-bond states,   has been demonstrated even in the case of 3D systems \cite{SBS1}.

In a recent paper we have introduced a  class of variational states, namely entangled-plaquette states (EPS), which provides an accurate ansatz to simulate the GS properties of quantum  spin models \cite{eps}. 
We have shown \cite{eps} that the EPS trial wave function (WF) allows to describe efficiently the GS of lattice Hamiltonians taking advantage of a simple variational Monte Carlo algorithm. Extremely accurate results have been obtained by us \cite{eps} and other investigators \cite{cps, cps1} for several frustrated (fermionic) and unfrustrated models.  

In this paper we study, as a paradigmatic example of frustrated system, the spin-$\frac{1}{2}$ AHM on the triangular lattice. Previous works, carried out by means of both QMC approaches  with a fixed-node approximation and variational calculations based on different classes of states, 
which explicitly incorporate antiferromagnetic  $\frac{2\pi}{3}$ N\`{e}el order, found values of the energy per site extrapolated to the thermodynamic limit between  $\sim -0.53$ and $\sim-0.5470$ \cite{huse, massimo, sorella, giamarchi, sorella1, paramekanti}.
Evidence of  long-range order  has been generally found with important differences, however, in the estimate of the order parameter. Our work,  is motivated, therefore, not only by the interesting physics which can emerge from the GS of a quantum antiferromagnet but also by methodological aspects: it is in fact important,  in order to asses its applicability to generic quantum spin models,  to benchmark the general  EPS ansatz    against other variational choices specifically designed to tackle a particular problem.
  
We carried out computer simulations on systems of size up to $N=324$ spins (previous applications of EPS on frustrated systems have been  performed on much smaller system sizes, up to $N=64$) \cite{eps}. We found for $N=36$, in the case of the XX model, an error on our estimated energy relative to the exact one \cite{runge} as small as $6\times 10^{-4}$ and an energy per site extrapolated to the thermodynamic limit of -0.4074(1).  At the Heisenberg point we obtained $E^{EPS}_0(N=36)=-0.55420(4)$,  which differs from the exact value by $\sim 1\%$, and  $E^{EPS}_0(\infty)=-0.5458(2)$, with a value for the order parameter of 0.3178(5).
\section{The EPS wave function}

In this section we describe the entangled-plaquette ansatz offering also some detail concerning the procedure adopted to minimize the energy and compute  GS physical observables. 
Given a collection of $N$ spins $\frac{1}{2}$ arranged on a lattice, a generic trial  WF can be expressed as:

\begin{equation}
|\psi\rangle=\sum_{\mathbf{n}}W(\mathbf{n})|\mathbf{n } \rangle
\end{equation}
where the $W(\mathbf{n})$'s are complex weights, $\mathbf{n} \rangle = |n_1,n_2, \ldots , n_N\rangle$ and $n_i =  \pm 1$ $ \forall$  $i=1, 2, \ldots,  N$ is the eigenvalue of $\sigma^z_i$. The variational principle ensures that the energy expectation value on a trial state is a rigorous upper bound of the GS energy, therefore, such a quantity can be evaluated by minimizing the trial energy with respect to the weights. To achieve this task one has to specify the form of $W(\mathbf{n})$, which corresponds to the choice of a particular variational ansatz. A desirable ansatz is one able to capture the GS physics of a system by means of few variational parameters, which can be efficiently optimized  on a computer. The entangled-plaquette ansatz is based on the following idea:

i) Cover the systems with $M$ plaquettes in such a way that $l$ sites labelled by $n_{1,P},n_{2,P},\ldots,n_{l,P}=\n_P$  belong to the $P_{th}$ plaquette.

 ii) Express the weight of a global spin configuration as a product of variational coefficients $C_1^{\n_1},C_2^{\n_2},\ldots,C_M^{\n_M}$   in biunivocal correspondence to the particular spin (e.g., along the $z$ axis) configuration of the plaquettes. Hence:

 \begin{equation}
\langle \n | \psi \rangle=W(\n)=\prod_{P=1}^M C_P^{\n_P}.
\label{coeff}
\end{equation}

In order to gain additional insight into the EPS ansatz let us discuss few exemplificative cases:

a) Each site is a plaquette: the wave function is  of the mean field form (i.e., correlations are neglected).

b) Plaquettes comprise more than one site and each site belongs to a single plaquette: with this choice it is possible to obtain reasonable estimates of the GS energy (up to corrections which scale with the plaquette boundaries) and short-range correlations (of the order of the plaquette size). 

c) Overlapping (i.e., entangled) plaquettes are used: this  ansatz clearly produces, accounting for the entangled nature of a quantum GS,   estimates of the energy and long-range correlations much more accurate than those computable with a WF whose weights are the product of non overlapping plaquettes \cite{duke, duke1} as in a) and b).

A variety of variational ansatze, as illustrated in  \cite{cps}, have a straigthforward representation in terms of EPS. This is, for example, the case of the WF introduced by Huse and Elser \cite{huse} for the AHM on the triangular lattice which contains up to three-spin correlations and a phase factor that depends on the system geometry. The accuracy of this variational choice can promptly be improved via a general (i.e., with no specific phase factor) EPS WF based on larger or differently shaped plaquettes, hence, including additional correlations. 
 In our calculations we start from the trial state described in a)  (i.e., $N=M$ and $l=1$) where  the weights and the initial spin configuration in the computational basis are randomly selected; we estimate, via Monte Carlo sampling, the expectation value of the energy as well as its derivatives with respect to the  $C_P^{\n_P}$'s; the value of these coefficients is then optimized by updating all of them of a small step against the gradient direction. Once convergence is reached the size of each plaquette is increased, naturally introducing correlations in the ansatz, and the procedure iterated. Finally, obtained the optimal GS WF, the mean value  of observables  other than the energy can be evaluated  by means of the Monte Carlo method as well.
An exhaustive  discussion on  the  particular suitablility of EPS to efficiently perform  the minimization procedure previously outlined is reported in  \cite{eps}.
It is important to mention that  the entangled-plaquette ansatz, as well as the proposed numerical protocol, is applicable to systems in any spatial dimension; moreover our method, as any purely variational (i.e., not based on imaginary time projection techniques) one, is sign-problem free allowing for the characterization of both unfrtustrated  and frustrated systems.

\section{Results}
In the following, we show results obtained with EPS for the 2D AHM on the triangular lattice. Computer simulations have been performed on several system sizes comprising up to $N=324$ sites. Our numerical findings are compared with exact diagonalization (ED) results available for $N=36$ \cite{runge} and estimates obtained with alternative computational approaches or variational ansatze \cite{huse, massimo, sorella, giamarchi, sorella1, paramekanti}. The AHM Hamiltonian is:
\begin{equation}
H= J^{xy}\sum_{<i,j>}({S}_i^x {S}_j^x+{S}_i^y {S}_j^y) +J^{z}\sum_{<i,j>}{S}_i^z {S}_j^z
\label{eq:heis}
\end{equation}
where $J^{xy}=J^z=1$, $S_i^{\alpha}$  is the spin-$\frac{1}{2}$ operator in the $\alpha$ direction acting on site $i$ and the  summations run over nearest-neighbour sites. The calculations were carried on in the $S^z_{TOT}=\sum_i^NS^z_i=0$ sector, being the total magnetization along  $z$ a good quantum number. Therefore, during the simulation, we update the system configuration proposing a spin-flip for couples of spins with opposite $S^z$. As a first benchmark we estimate the GS energy of the XX model (i.e., $J^z=0$, $J^{xy}=1$) with PBC. Numerical results, shown in table \ref{tab:1}, refer to different size $l$ of the  plaquettes. For $l=4$ the relative error  of the EPS energy of the $N=36$ system with respect to the exact result is smaller than $2\%$, and decreases down to $6.5\times 10^{-4}$ when $l=16$. Our estimate, $E_0^{EPS}(N=36)=-0.41069(3)$,  compares favorably to those obtained with alternative trial WF's \cite{sorella, paramekanti}.  For example the trial state introduced  in \cite{huse}, usually employed as a starting point for fixed-node based QMC calculations, leads to a value of the energy per site of about $-0.4039$ \cite{sorella} (i.e., approximately $1.7\%$ higher than the exact result).  

\begin{table}[b]
\caption{\label{tab:1} Variational GS energy per site (in units of  $J^{xy}$) computed, using the entangled-plaquette ansatz, for the spin-$\frac{1}{2}$ antiferromagnetic XX model on a triangular lattice of size $N$ with PBC. Error bars are in parenthesis; the number of spins per plaquette is reported in the first column. The exact result for $N=36$ is also shown for comparison \cite{runge}.}
\centering
\begin{tabular}{@{}cccc}
\br
$l$ & N=36 & N=144 & N=324 \\
\mr
4 & -0.40426(5) & -0.39954(5) & -0.39875(5) \\
6 & -0.40626(4) & -0.40197(4) & -0.40120(4) \\
9 & -0.40956(3) & -0.40609(3) & -0.40545(3) \\
12 & -0.41006(3) & -0.40685(3) & -0.40630(3) \\
16 & -0.41069(3) & -0.40793(3) & -0.40737(3) \\
 Exact & -0.410957 & - & - \\
 \br
\end{tabular}
\end{table}

\begin{figure}[t]
\centerline{\includegraphics[scale=0.7]{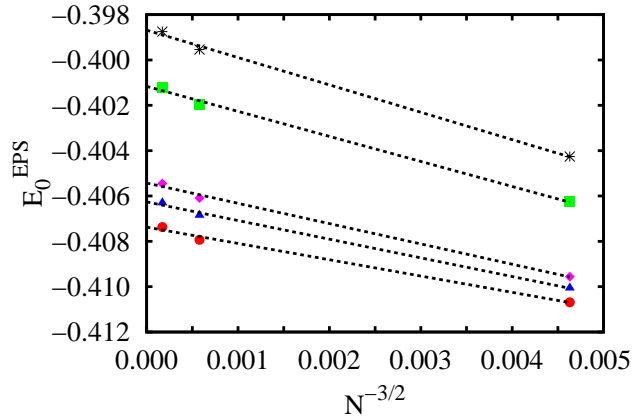}}
\caption{Size dependence of the GS energy per site for the spin-$\frac{1}{2}$ antiferromagnetic XX model on the triangular lattice with PBC. Different symbols refer to different plaquette size $l$ (i.e., stars: $l=4$, boxes: $l=6$, diamonds: $l=9$, triangles: $l=12$, circles: $l=16$. Error bars, not shown for clarity, are smaller than the symbol size. Each dashed line is a fit to numerical data with the same $l$ (see text).}
\label{fig:1}
\end{figure}

The extrapolation to the infinite lattice limit of the energies in table \ref{tab:1} has been performed by assuming the functional scaling  $E_0^{EPS}(N)=E_0^{EPS}(\infty)+\frac{\beta}{N^{\frac{3}{2}}}$;
 the dependence of the GS energy per site on the system size, as well as the fitting functions for different values of $l$, are reported in figure \ref{fig:1}. Extrapolated energy values range from -0.3987(2) ($l=4$) to -0.4074(1) ($l=16$) and are indistinguishable, taking into account the statistical uncertainty, from the results obtained when $N=324$. For this size, therefore, the  thermodynamic limit appears fully recovered. The EPS extrapolated energy is slightly lower than that computed by extrapolating exact results up to $N=36$ (i.e., $E_0^{ED}(N=\infty)=-0.4066$), illustrating the importance of studying system sizes larger than those treatable via ED techniques. 
 It has to be mentioned that the small deviation shown by our data in figure  \ref{fig:1}  from a linear  scaling, being  the same  regardless the size of the plaquettes, should not be due to a loss in accuracy of the EPS  WF as the system size increases; more plausibly, it  can be ascribed to higher order corrections in the series expansion of the energy as a function of $\frac{1}{N}$.  Moreover  the energy estimates (for a fixed lattice size) do not show evidence of convergence with the plaquette sizes that we have been able to utilize in this work.  This is not surprising (in fact the EPS ansatz is formally exact  when the plaquette size is as big as that of the whole system) and simply means that the exact result, even if the energy differences for $l > 9$ are quite small, is not yet reached. As a consequence one can expect,  on further increasing the plaquette size, an improvement of the GS energy. It is however remarkable, taking into account the methodological aspect of our paper, that  our best variational upper bounds, obtained by using plaquettes of $16$ sites, already provide   extremely accurate energy estimates.

Next, we discuss our findings for the GS of the Hamiltonian (\ref{eq:heis}) at the Heisenberg point (i.e., $J^{xy}=J^z=1$). GS energies per site are reported in table \ref{tab:2}. The EPS GS energy computed for $N=36$ differs from the exact one \cite{runge} by $\sim 1\%$, while, calculations based on ordered WF's yield typically higher values of the GS energy between $\sim -0.5367$ \cite{huse} and $\sim -0.5515$ \cite{sorella1}. When $N$ increases, the general EPS WF remains competitive with respect to  accurate  choices specifically adapted to the  triangular lattice geometry. Particularly, our energies for $N=144$ and $N=324$ differ from the best (to our knowledge) variational estimates \cite{sorella1} by less than $0.25\%$.  

\begin{table}[t]
\caption{\label{tab:2} Variational GS energy per site (in units of  $J^{xy}$) computed, by using the entangled-plaquette ansatz, for the spin-$\frac{1}{2}$ AHM on a triangular lattice of size $N$ with PBC. Error bars are in parenthesis; $N$ plaquettes comprising $16$ sites have been employed. The exact result for $N=36$ is also shown for comparison \cite{runge}.}
\centering
\begin{tabular}{@{}ccc}
\br
$N$ & $E^{EPS}_0$  & Exact \\
\mr
36 & -0.55420(5) & -0.560373  \\
144 & -0.54700(8) & - \\
324 & -0.5459(1) & - \\
\br
\end{tabular}
\end{table}

\begin{figure}[b]
\centerline{\includegraphics[scale=0.7]{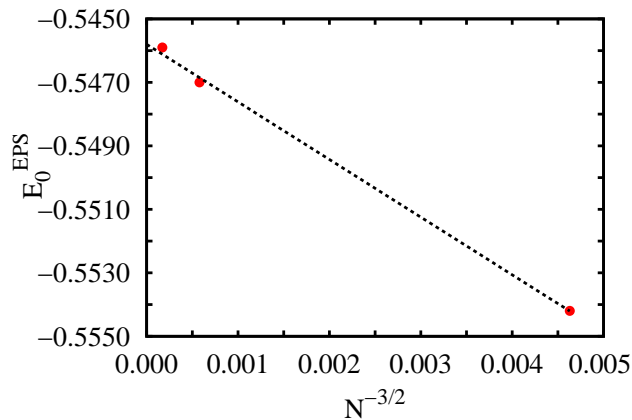}}
\caption{Size dependence of the GS energy per site of the spin-$\frac{1}{2}$ AHM on the triangular lattice with PBC. Error bars, not shown for clarity, are smaller than the symbol size. The dashed line is a fit to numerical data.}
\label{fig:2}
\end{figure}

As a result of a finite-size scaling analysis of the EPS energies per site carried out by extrapolating the numerical data in table \ref{tab:2} (see figure \ref{fig:2}) we find $E_0^{EPS}(\infty)=-0.5458(2)$, which is in excellent agreement with the most accurate results derived in previous works \cite{sorella,sorella1}.

GS estimates of the spin-spin correlation function $Corr^{EPS}_0(\mathbf{r}_{MAX},N)=\langle\mathbf{S_0}\cdot\mathbf{S_{r_{MAX}}}\rangle_{EPS}$ computed at the maximum distance (taking into account the PBC) on the lattice are reported in table \ref{tab:3}.  The magnetic order parameter of the system defined as $m^{2}=lim_{N\rightarrow\infty}Corr^{}_0(\mathbf{r}_{MAX},N)$ is a quantity of central importance, being directly related to the presence of long-range order. Wether  or not the spin-$\frac{1}{2}$ AHM displays long-range order has been a long lasting-question, however,  the positive answer, on the basis of analytical \cite{miy, zheng} and numerical \cite{huse, massimo, sorella, giamarchi, sorella1, runge, white, ron} results, is commonly accepted. Estimates of the  order parameter expressed as a fraction of its classical value  range from  $\sim0.4$ \cite{sorella, white} to $\sim0.72$ \cite{giamarchi}. 
Our computed spin-spin correlations for $N=144$ and $N=324$ are indistinguishable (within the statistical uncertainty). Hence we can consistently assume $m^{EPS}=(Corr^{EPS}_0(\mathbf{r}_{MAX},N=324))^{\frac{1}{2}}=0.3178(5)$ (i.e., $\sim 64\%$ of the classical result). This value  of the order parameter, approximately 2\% lower than that obtained with the same numerical approach for the AHM on the square lattice  \cite{eps}, is consistent with the long-range ordered nature of the system found in previous numerical studies  \cite{sorella, zheng, white, massimo, giamarchi}. 

\begin{table}[t]
\caption{\label{tab:3} Variational GS spin-spin correlation   (in units of  $J^{xy}$), at the maximum distance (with PBC) on the lattice.  Error bars are in parenthesis; $N$ plaquettes comprising $16$ sites have been employed.}
\centering
\begin{tabular}{@{}cc}
\br
$N$ & $Corr^{EPS}_0(\mathbf{r}_{MAX},N)$   \\
\mr
36 & 0.1294(1)   \\ 
144 & 0.1011(2)  \\ 
324 & 0.1010(3)  \\
\br
\end{tabular}
\end{table}

\section{Conclusions}

The general entangled-plaquette wave function has been employed to study the ground-state properties of the spin-$\frac{1}{2}$ antiferromagnetic XX and Heisenberg Hamiltonian on the triangular lattice. With this ansatz and a simple variational Monte Carlo algorithm for the energy minimization, we found extremely accurate  energy estimates for finite systems, as well as, by extrapolating,  in the thermodynamic limit. The accuracy of our results compares  well with that provided by the best alternative methods or variational ansatze, which, when possible, (as in the case of the triangular lattice), are adjusted, to improve the estimates, by incorporating physical features of the particular system of interest.
It is crucial to stress that the  trial state utilized in this work, although not  including any problem dependent term, provides  a reasonable  estimate even  of the system order parameter; therefore, the variational class of entangled-plaquette states   can be regarded as a reliable option to describe the GS of general frustrated (i.e., not treatable with standard quantum Monte Carlo methods) physical models  on a lattice with no need of any a-priori knowledge.
\ack
We acknowledge discussions with M. Boninsegni and D. Yudin. This work has been supported by the DFG (FOR 635) and the EU project QUEVADIS.


\end{document}